\documentclass{natureprintstyle}

\usepackage{graphicx}
\usepackage{amssymb}

\newcommand{\apj}{Astrophys. J.}           
\newcommand{\mnras}{Mon. Not. R. Astron. Soc.}       
\newcommand{\nat}{Nature}
\newcommand{\aap}{Astron. Astrophys.}
\newcommand{\araa}{Annual Rev. Astron. Astrophys.}

\newcommand{\pasp}{Pubbl. Astron. Soc. Pacific}

\newcommand{\sauron}{\texttt{SAURON}}
\newcommand{\atl}{ATLAS$^{\rm 3D}$}
\newcommand{\kms}{\hbox{km s$^{-1}$}}
\newcommand{\re}{\hbox{$R_{\rm e}$}}

\title{A systematic variation of the stellar initial mass function in early-type galaxies}

\author{Michele Cappellari$^1$,
Richard M. McDermid$^{2}$,
Katherine Alatalo$^3$,
Leo Blitz$^3$,
Maxime Bois$^{4}$,
Fr\'ed\'eric Bournaud$^{5}$,
M.~Bureau$^1$,
Alison F. Crocker$^6$,
Roger L. Davies$^1$,
Timothy A. Davis$^{1,7}$,
P. T. de Zeeuw$^{7,8}$,
Pierre-Alain Duc$^{5}$,
Eric Emsellem$^{7,9}$,
Sadegh Khochfar$^{10}$,
Davor Krajnovi\'c$^7$,
Harald Kuntschner$^{7}$,
Pierre-Yves Lablanche$^{7,9}$,
Raffaella Morganti$^{11,12}$,
Thorsten Naab$^{13}$,
Tom Oosterloo$^{11,12}$,
Marc Sarzi$^{14}$,
Nicholas Scott$^{1,15}$,
Paolo Serra$^{11}$,
Anne-Marie Weijmans$^{16}$
\& Lisa M. Young$^{17}$
}

\begin{document}

\maketitle

\begin{abstract}
Much of our knowledge of galaxies comes from analysing the radiation emitted by their stars. It depends on the stellar initial mass function (IMF) describing the distribution of stellar masses when the population formed. Consequently knowledge of the IMF is critical to virtually every aspect of galaxy evolution. More than half a century after the first IMF determination\cite{Salpeter1955}, no consensus has emerged on whether it is universal in different galaxies\cite{Bastian2010}. Previous studies indicated that the IMF and the dark matter fraction in galaxy centres cannot be both universal\cite{Cappellari2006,Tortora2009,Treu2010,Thomas2011,Dutton2011imf}, but they could not break the degeneracy between the two  effects. Only recently indications were found that massive elliptical galaxies may not have the same IMF as our Milky Way\cite{vanDokkum2010}. Here we report unambiguous evidence for a strong systematic variation of the IMF in early-type galaxies as a function of their stellar mass-to-light ratio, producing differences up to a factor of three in mass. This was inferred from detailed dynamical models of the two-dimensional stellar kinematics for the large \atl\ representative sample\cite{Cappellari2011a} of nearby early-type galaxies spanning two orders of magnitude in stellar mass. Our finding indicates that the IMF depends intimately on a galaxy's formation history.
\end{abstract}

\begin{figure*}
\begin{center}
\includegraphics[width=0.75\textwidth]{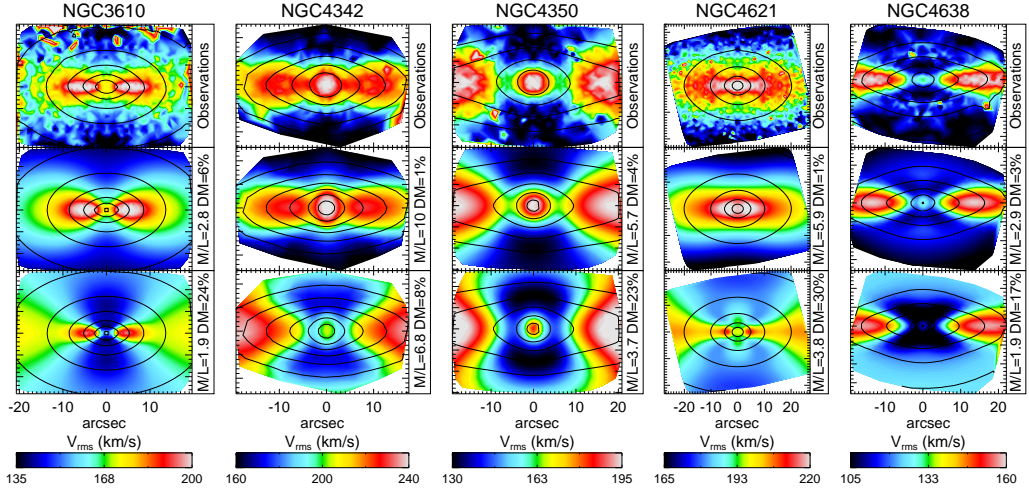}
\end{center}
\caption{{\bf Disentangling the stellar and dark mass with integral-field stellar kinematics.} The top panels show the symmetrized \sauron\ stellar kinematics $V_{\rm rms}=\sqrt{V^2+\sigma^2}$ for five galaxies representing a variety of shapes of the kinematics fields, and spanning a range in $(M/L)_{\rm stars}$ values. Here $V$ is the mean stellar velocity and $\sigma$ is the stellar velocity dispersion. The middle panel is the best-fitting dynamical model\cite{Cappellari2008} with a standard\cite{navarro96} dark halo (model {\bf b} in Table~1). The bottom panel is a dynamical model where the $(M/L)_{\rm stars}$ was fixed to be 0.65 times the best-fitting one. Where this decrease in $(M/L)_{\rm stars}$ represents the change in mass between a Salpeter and Kroupa IMF.  The other three model parameters, the galaxy inclination $i$, the orbital anisotropy $\beta_z$ and the halo total mass $M_{200}$, were optimized to fit the data, but cannot provide an acceptable description of the observations. This plots shows that, for a standard halo profile, the data tightly constrain both the dark matter fraction and $(M/L)_{\rm stars}$. The effect would be even more dramatic if we had assumed a more shallow inner halo profile. The contours show the observed and modelled surface brightness respectively. The values of $(M/L)_{\rm stars}$ and the fraction of dark matter within a sphere with radius equal to the projected half-light radius \re\ are printed next to each panel. The galaxy names are given at the top.}
\end{figure*}

\begin{figure*}
\begin{center}
\includegraphics[width=0.75\textwidth]{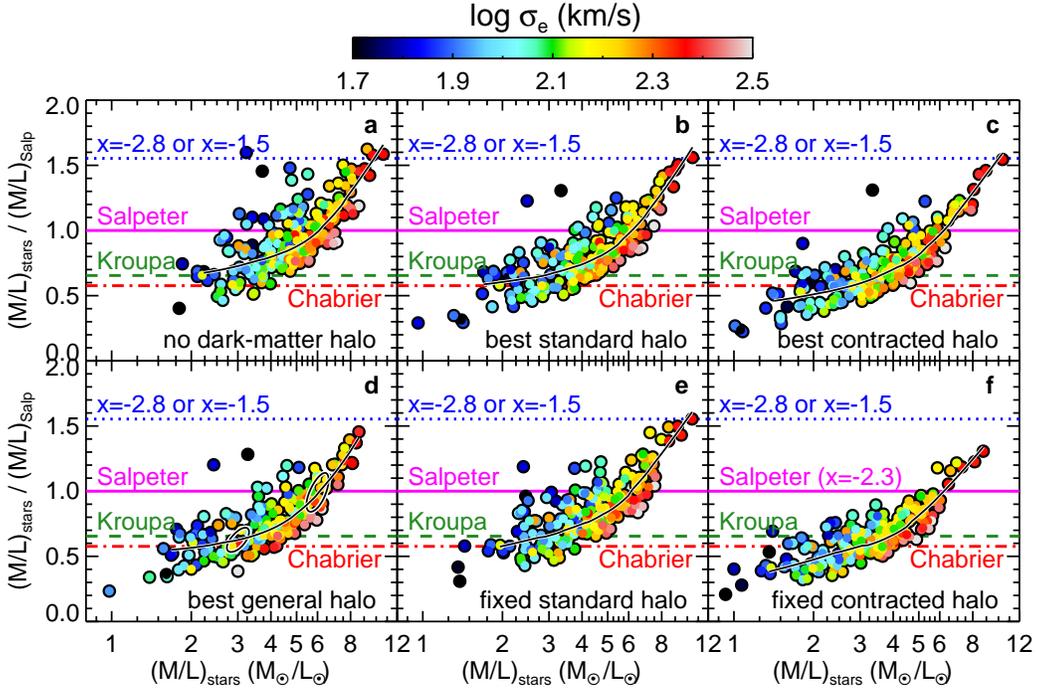}
\end{center}
\caption{{\bf The systematic variation of the IMF in early-type galaxies.} The six panels show the ratio between the $(M/L)_{\rm stars}$ of the stellar component, determined using dynamical models, and the $(M/L)_{\rm Salp}$ of the stellar population, measured via stellar population models with a Salpeter IMF, as a function of $(M/L)_{\rm stars}$. The black solid line is a {\em loess} smoothed version of the data. Colours indicate the galaxies' stellar velocity dispersion $\sigma_{\rm e}$, which is related to the galaxy mass. The horizontal lines indicate the expected values for the ratio if the galaxy had  (i) a Chabrier IMF (red dash-dotted line); (ii) a Kroupa IMF (green dashed line); (iii) a Salpeter IMF ($x=-2.3$, solid magenta line) and two additional power-law IMFs with (iv) $x=-2.8$ and (v) $x=-1.5$ respectively (blue dotted line).
Different panels correspond to different assumptions for the dark matter halos employed in the dynamical models as written in the black titles. Details about the six sets of models are given in Table~1. A clear curved relation is visible in all panels. Panels {\bf a}, {\bf b} and {\bf e} look quite similar, as for all of them the dark matter contributes only a small fraction (zero in {\bf a} and a median of 12\% in {\bf b} and {\bf e}) of the total mass inside a sphere with the projected size of the region where we have kinematics (about one projected half-light radius \re). Panel {\bf f} with a fixed contracted halo, still shows the same IMF variation, but it is almost systematically lower by 35\% in $(M/L)_{\rm stars}$ reflecting the increase in dark matter fraction. The two black thick ellipses plotted on top of the smooth relation in panel {\bf d} show the representative 1$\sigma$ error for one measurement at the given locations. We excluded from the plot the galaxies with very young stellar population (selected as having H$\beta>2.3$ \AA\ absorption). These galaxies have strong radial gradients in their population, which break our assumption of spatially constant $M/L$ and makes both $(M/L)_{\rm Salp}$ and $(M/L)_{\rm stars}$ inaccurate. }
\end{figure*}

As part of the \atl\ project\cite{Cappellari2011a}, we obtained integral-field maps of stellar kinematics for a volume-limited sample of 260 early-type (elliptical and lenticular) galaxies. They were selected to be closer than 42 Mpc and to have a $K_s$-band total magnitude $M_K<-21.5$ mag ($M\gtrsim 6\times10^9$ M$_\odot$), as determined from the Two Micron All Sky Survey (2MASS) at our adopted distances. Homogeneous imaging for all the galaxies in the $r$-band was obtained in major part from the Sloan Digital Sky Survey (SDSS) DR8 and completed with our own photometry.
\let\thefootnote\relax\footnote{
\begin{affiliations}
 \item Sub-department of Astrophysics, Department of Physics, University of Oxford, Denys Wilkinson Building, Keble Road, Oxford OX1 3RH
\item Gemini Observatory, Northern Operations Centre, 670 N. A`ohoku Place, Hilo, HI 96720, USA
\item Department of Astronomy, Campbell Hall, University of California, Berkeley, CA 94720, USA
\item Observatoire de Paris, LERMA and CNRS, 61 Av. de l'Observatoire, F-75014 Paris, France
\item Laboratoire AIM Paris-Saclay, CEA/IRFU/SAp – CNRS – Universit\'e Paris Diderot, 91191 Gif-sur-Yvette Cedex, France
\item Department of Astrophysics, University of Massachusetts, 710 North Pleasant Street, Amherst, MA 01003, USA
\item European Southern Observatory, Karl-Schwarzschild-Str. 2, 85748 Garching, Germany
\item Sterrewacht Leiden, Leiden University, Postbus 9513, 2300 RA Leiden, the Netherlands
\item Universit\'e Lyon 1, Observatoire de Lyon, Centre de Recherche Astrophysique de Lyon and Ecole Normale Sup\'erieure de Lyon, 9 avenue Charles Andr\'e, F-69230 Saint-Genis Laval, France
\item Max-Planck Institut f\"ur extraterrestrische Physik, PO Box 1312, D-85478 Garching, Germany
\item Netherlands Institute for Radio Astronomy (ASTRON), Postbus 2, 7990 AA Dwingeloo, The Netherlands
\item Kapteyn Astronomical Institute, University of Groningen, Postbus 800, 9700 AV Groningen, The Netherlands
\item Max-Planck Institut f\"ur Astrophysik, Karl-Schwarzschild-Str. 1, 85741 Garching, Germany
\item Centre for Astrophysics Research, University of Hertfordshire, Hatfield, Herts AL1 9AB, UK
\item Centre for Astrophysics \& Supercomputing, Swinburne University of Technology, PO Box 218, Hawthorn, VIC 3122, Australia
\item Dunlap Institute for Astronomy \& Astrophysics, University of Toronto, 50 St. George Street, Toronto, ON M5S 3H4, Canada
\item Physics Department, New Mexico Institute of Mining and Technology, Socorro, NM 87801, USA
\end{affiliations}
}

\begin{table*}
\centering
\caption{{\bf The axisymmetric dynamical models}}
\begin{tabular}{cp{8.5cm}c}
\hline
Panel in Fig.~2 & Description of the model & Fitted model parameters \\
\hline
{\bf a} & Galaxy model in which the total mass traces the observed galaxy light distribution. Any dark matter, if present, follows the stellar distribution. & $i,\beta_z,(M/L)_{\rm total}$ \\
{\bf b} & Galaxy stellar component embedded in a spherical standard dark matter halo\cite{navarro96} with inner density $\rho(r)\propto r^{-1}$ for radii $r\ll r_s$ and outer density $\rho(r)\propto r^{-3}$ for $r\gg r_s$. The halo total mass $M_{200}$ is fitted, while $r_s$ is uniquely specified\cite{Klypin2011} by $M_{200}$. & $i,\beta_z,(M/L)_{\rm stars},M_{200}$ \\
{\bf c} & Model with a standard\cite{navarro96} halo contracted\cite{Gnedin2011} according to the observed galaxy stellar density. The halo mass is fitted, while $r_s$ is specified\cite{Klypin2011} by $M_{200}$. & $i,\beta_z,(M/L)_{\rm stars},M_{200}$ \\
{\bf d} & Model with a general halo inner density $\rho(r)\propto r^\gamma$ with fitted slope ($-1.6<\gamma<0$) and total mass. The outer density becomes $\rho(r)\propto r^{-3}$ as in the standard halo\cite{navarro96} at radii $r\gg r_s=20$ kpc. & $i,\beta_z,(M/L)_{\rm stars},\gamma,M_{200}$ \\
{\bf e} & Model with a fixed standard halo\cite{navarro96} with $M_{200}$ specified\cite{Moster2010} by the measured galaxy stellar mass and $r_s$ specified\cite{Klypin2011} by $M_{200}$. & $i,\beta_z,(M/L)_{\rm stars}$ \\
{\bf f} & Model with a fixed standard halo\cite{navarro96} contracted\cite{Gnedin2011} according to the observed galaxy stellar density. $M_{200}$ is specified\cite{Moster2010} by the measured galaxy stellar mass and $r_s$ is specified\cite{Klypin2011} by $M_{200}$. & $i,\beta_z,(M/L)_{\rm stars}$ \\
\hline
\end{tabular}
\end{table*}

For all galaxies, we constructed six sets of dynamical models\cite{Cappellari2008}, which include an axisymmetric stellar component and a spherical dark halo, and fit the details of both the projected stellar distribution\cite{Emsellem1994} and the two-dimensional stellar kinematics\cite{Cappellari2011a} (Fig.~1). While the shape of the stellar component can be inferred directly from the galaxy images, the  dark halo shape must be a free parameter of the models. We explored with the models a variety of plausible assumptions for the halo to test how these can affect our result. Our halo models include as limiting cases a maximum-ignorance one, where the halo parameters are directly fitted to the stellar kinematics, and some completely fixed ones, where the halo follows the predictions of numerical simulations\cite{navarro96,Klypin2011,Gnedin2011}. A detailed description of the model parameters is provided in Table~1. The key parameter we extract from all the models is the ratio  $(M/L)_{\rm stars}$ between the luminosity (in the $r$-band) and the mass of the stellar component. As illustrated in Fig.~1, the availability of integral-field data is the key to accurately separate the stellar mass from the possible dark matter using dynamical models and determine $(M/L)_{\rm stars}$. In fact changes in $(M/L)_{\rm stars}$ at the level expected for IMF variations cause dramatic changes to the quality of the model fits.

We also measured the $(M/L)_{\rm pop}$ of the stellar population by fitting\cite{Cappellari2004} the observed spectra using a linear combination of single stellar population synthetic spectra\cite{Vazdekis2010} of different ages ($t$) and metallicities ($[M/H]$), adopting for reference a Salpeter\cite{Salpeter1955} IMF ($\xi(m)\propto m^x=m^{-2.3}$). The models adopt standard lower and upper mass cut-offs for the IMF of 0.1 and 100 M$_\odot$, respectively.  We used linear regularization to reduce noise and produce smooth $M(t,[M/H])$ solutions consistent with the observations. The resulting $(M/L)_{\rm pop}$ is that of the composite stellar population, and excludes the gas lost during stellar evolution. If all this gas was retained in the galaxies in gaseous form, it would systematically increase $(M/L)_{\rm pop}$ by about 30\%\cite{Maraston2005}. However most of it is likely recycled into stars or expelled to larger radii. Although the results are cleaner using our full spectrum fitting approach\cite{Cappellari2004}, similar conclusions are reached when the galaxies are approximated as one single stellar population, or when $(M/L)_{\rm pop}$ is computed using different population codes\cite{Vazdekis2010,Maraston2005,bruzual03} and with a more traditional approach which only focuses on the strength of a few stellar absorption spectral lines. Systematic offsets of about 10\% in $(M/L)_{\rm pop}$ exist between the predictions of different population models, for an identical set of assumed population parameters, with the adopted one being in the middle of the others. This sets the uncertainty in the absolute normalization of our plots. The random errors for our adopted population code\cite{Vazdekis2010} were estimated by applying the same spectral fitting approach to our integral-field spectroscopy data and to independent spectra obtained for a subset of 57 of our galaxies by the SDSS survey. We inferred an rms scatter of 12\% in each individual $(M/L)_{\rm pop}$ determination.

The ratio between the dynamically-derived $(M/L)_{\rm stars}$ and the population-derived $(M/L)_{\rm Salp}$, using a fixed Salpeter IMF, is shown in Fig.~2 as a function of $(M/L)_{\rm stars}$. We compare the observed ratio with the expected one if the galaxy had the `light' Kroupa\cite{Kroupa2001} or Chabrier\cite{Chabrier2003} IMFs, which are similarly deficient in low mass stars; the `standard' Salpeter IMF, which is described by a simple power-law in stellar mass with exponent $x=-2.3$; and two additional `heavy' power-law IMFs with $x=-2.8$ and $x=-1.5$ respectively. The last two IMFs predict the same $(M/L)_{\rm pop}$. But while in the first case the stellar population is dominated by dwarf stars, in the second case the large $(M/L)_{\rm pop}$ is due to stellar remnants: black holes, neutron stars and white dwarfs. The dynamical mass  measurements do not constrain the shape of the IMF directly, but only the overall mass normalization, and for this reason cannot distinguish between the two cases.

The results from  all sets of dynamical models are  consistent with a similar systematic variation of the IMF normalization, by up to a factor of three in mass. A clear trend is visible in particular for the most general of our set of models (panel {\bf d} of Fig.~2), which makes virtually no assumptions on the halo shape but fits it directly to the data. However similar trends are visible for all our plausible assumptions for the dark halo mass and profile as predicted by numerical simulations. This shows that, although our result does not depend on the correctness of the assumed halo model, it is entirely consistent with standard model predictions for the halo. For increasing $(M/L)_{\rm stars}$ the normalization of the inferred IMF varies from the one of Kroupa/Chabrier up to an IMF more massive than Salpeter.  The trend in IMF is still clearly visible when selecting a subset of 60 galaxies lying outside the Virgo galaxy cluster, with the most accurate distances and the best models fits. This shows that it cannot be due to biases in the models, distances, or to effects related to the cluster environment. The knee in the relation at $(M/L)_{\rm stars}\approx6$ ($r$-band) shows that the lowest $(M/L)_{\rm stars}$ values mainly reflect the age and metallicity of the population (with younger ages or lower [M/H] decreasing $(M/L)_{\rm pop}$), while the largest $(M/L)_{\rm stars}$ values mainly reflect their dwarf or remnants dominated IMF. The models with contracted halos show the same IMF trend as the other models. However contracted halos predict too little stellar mass for many of the galaxies with low $(M/L)_{\rm stars}$, even for the `lightest' Kroupa/Chabrier IMF. This suggests that contraction may not happen in most real galaxies, in agreement with recent numerical simulations\cite{Duffy2010}.

Our result reconciles a number of apparently contradictory results on the normalization of the IMF accumulated in the past decade. The Kroupa/Chabrier-like normalization at low  $(M/L)_{\rm stars}$ agrees with the one inferred for spiral galaxies\cite{Bell2001}. The \atl\ project discovered that early-type galaxies with the lowest  $(M/L)_{\rm pop}$ resemble spiral galaxies with their gas and dust removed\cite{Cappellari2011b} and thus a similarity of IMF should be expected. The Kroupa/Chabrier normalization is also consistent with previous findings that this normalization is required not to over-predict the mass of early-type galaxies as a class\cite{Cappellari2006,Renzini2006,Ferreras2008}. A Salpeter normalization at larger $(M/L)_{\rm stars}$ is consistent on average with results from strong gravitational lensing\cite{Treu2010}, which are restricted to the galaxies with the largest velocity dispersions ($\sigma\gtrsim200$ \kms). Finally, the normalization more massive than Salpeter for some of the galaxies with large $(M/L)_{\rm stars}$ is broadly consistent with the finding from the depth of spectral features of eight massive galaxies\cite{vanDokkum2010} that indicate they must be dominated by a population of dwarf stars.

If instead the largest $(M/L)_{\rm pop}$ were due to stellar remnants, our results would be consistent with indirect arguments based on the relation between the colour of a stellar population and its fraction of ionizing photons, suggesting an IMF slope becoming flatter for more massive and star forming galaxies\cite{Hoversten2008,Gunawardhana2011}. However our result is difficult to compare directly, due to the large difference in the sample selections. Moreover these studies\cite{Hoversten2008,Gunawardhana2011} measure the instantaneous IMF, when the stars are forming, while all previous studies we mentioned, including the one in this Letter, measure the `integrated' galaxy IMF (IGIMF) resulting from the cumulative history of star formation\cite{Kroupa2003} and evolutionary mechanisms that the galaxy has experienced.

The discovered trend in IMF is also consistent with previous findings that the {\em total} $M/L$ in the centre of galaxies varies by at least a factor of two more than one would expect for a stellar population with constant dark matter fraction and a universal IMF\cite{Cappellari2006}. But various previous attempts could not distinguish whether the mass discrepancy was due to non-universality of dark matter or IMF\cite{Tortora2009,Treu2010,Dutton2011imf,Thomas2011,Deason2011dm}. The studies were limited either by small samples or non optimal data\cite{Cappellari2006,Thomas2011} or employed simplified galaxy models that could bias the quantitative interpretation of the results\cite{Tortora2009,Treu2010,Dutton2011imf,Deason2011dm}. We finally resolve both of these limitations.

Our study convincingly demonstrates that the assumption of a universal IMF, which is still adopted in nearly every aspect of galactic astrophysics, stellar populations and cosmology, is inconsistent with real galaxies. The results presented pose an interesting challenge to galaxy formation models, which will have to explain how stars `know' what kind of galaxy they will end up inside. A possible explanation would be for the IMF to depends on the prevailing physical conditions when the galaxy formed the bulk of their stars. Although galaxies merge hierarchically, there is growing evidence that present-day massive early-type galaxies formed most of their stars in more intense starbursts and at higher redshift than spiral galaxies. This could lead to the observed difference in IMF. Unfortunately there is no consensus in the theoretical models for how the IMF should vary with physical conditions.
A new generation of  theoretical and observational studies will have to provide insight into which physical mechanisms are responsible for the systematic IMF variation we find.\\

\noindent
\textbf{\textsf{\footnotesize Received 13 December 2011; accepted 13 February 2012.}}

\begin{addendum}
 \item[Acknowledgements] MC acknowledges support from a Royal Society University Research Fellowship.
This work was supported by the rolling grants `Astrophysics at Oxford' from the UK Research Councils. RLD acknowledges support from Christ Church, Oxford and from the Royal Society in the form of a Wolfson Merit Award.
SK acknowledges support from the the Royal Society Joint Projects Grant.
RMcD is supported by the Gemini Observatory.
TN and MBois acknowledge support from the DFG Cluster of Excellence `Origin and Structure of the Universe'.
MS acknowledges support from a STFC Advanced Fellowship.
NS and TD acknowledge support from an STFC studentship.
\item[Author Contributions] All authors contributed extensively to the work presented in this paper.
 \item[Author Information] Reprints and permissions information is available at
www.nature.com/reprints. The authors declare no competing financial interests.
Readers are welcome to comment on the online version of this article at
www.nature.com/nature. Correspondence and requests for materials should be addressed to M.C.~(cappellari@astro.ox.ac.uk).
\end{addendum}

\end{document}